\documentclass[12pt]{iopart}
\usepackage{graphicx}
\begin{document}

\title[A Study of Individuals Shifting Between Political Communities on Social Media]{Polarization Dynamics: A Study of Individuals Shifting Between Political Communities on Social Media}

\author{Federico Albanese\textsuperscript{1,*}, Esteban Feuerstein\textsuperscript{2} and Pablo Balenzuela\textsuperscript{3}}

\address{\textsuperscript{1} Instituto de Investigaci\'on en Ciencias de la Computaci\'on, CONICET- Universidad de Buenos Aires, Argentina}
\address{\textsuperscript{2} Departamento de Computaci\'on, Facultad de Ciencias Exactas y Naturales, Universidad de Buenos Aires, Argentina}
\address{\textsuperscript{3} Departamento de Física, Facultad de Ciencias Exactas y Naturales, Universidad de Buenos Aires, Argentina}
\address{\textsuperscript{*} Author to whom any correspondence should be addressed.}
\ead{falbanese@dc.uba.ar}
\vspace{10pt}

\begin{abstract}
Individuals engaging on social media often tend to establish online communities where interactions predominantly occur among like-minded peers. While considerable efforts have been devoted to studying and delineating these communities, there has been limited attention directed towards individuals who diverge from these patterns. In this study, we examine the community structure of re-post networks within the context of a polarized political environment at two different times. We specifically identify individuals who consistently switch between opposing communities and analyze the key features that distinguish them. Our investigation focuses on two crucial aspects of these users: the topological properties of their interactions and the political bias in the content of their posts. Our analysis is based on a dataset comprising 2 million tweets related to US President Donald Trump, coupled with data from over 100,000 individual user accounts spanning the 2020 US presidential election year. Our findings indicate that individuals who switch communities exhibit disparities compared to those who remain within the same communities, both in terms of the topological aspects of their interaction patterns (pagerank, degree, betweenness centrality.) and in the sentiment bias of their content towards Donald Trump.
\end{abstract}

%
\vspace{2pc}
\noindent{\it Keywords}: Polarization, Community detection, Sentiment Analysis, Social Media.
%
%
%
%

\section{Introduction}

Social media platforms have become influential spaces for political discourse, enabling users to express their opinions and engage with political figures. However, the exchange of ideas among users with opposing ideologies remains limited, primarily due to the formation of echo chambers \cite{jamieson2008echo}. Echo chambers refer to homogeneous communities where individuals gather based on their shared beliefs, limiting their exposure to alternative viewpoints and hindering meaningful debates \cite{aruguete2018time}. These communities have been found in multiple platforms \cite{cinelli2021echo,quattrociocchi2016echo} and contexts such as the 2010 U.S. congressional midterm elections \cite{conover2011political}, the climate change debate \cite{haussler2018heating} or the 2011 parliament elections in Germany \cite{dang2013investigation}.

Prior studies have extensively documented the adverse implications associated with closed communities and echo chambers. Specifically, these include the increase in negative discourse, hate speech and political extremism \cite{lima2018inside}, confirmation bias (i.e. the users tendency to seek out and receive information that strengthens their preferred narrative) \cite{quattrociocchi2016echo} and spreading of rumors and fake news \cite{del2016spreading,choi2020rumor}.

The exploration of how users change their communities constitutes a vital area of investigation within computational social science \cite{wang2020jump}. This phenomenon holds significant real-world implications, particularly as political elections can hinge upon individuals who swing from one political party to another \cite{mayer2008swing}. By studying the behavioral patterns of users within these dynamic social networks, valuable insights can be gained for the design of better communities. Additionally, the study of users who transition from a political community to other political community helps in the comprehension of polarization and depolarization processes, thus contributing to the development of strategies aimed at mitigating polarization and extreme ideological divisions, fostering a healthier political discourse.  

There is limited research on individuals who transition between ideological communities. In this context, Wang et al. focus on understanding the ``bandwagon fan'' phenomenon in Reddit sport communities, showing how some sports fans start following a new sports team solely because of its recent success \cite{wang2020jump}. Albanese et al. propose a method based on XGBoost, the Louvain method for community detection and topic modeling to detect users who change their political community on Twitter \cite{DBLP:conf/amw/AlbaneseFLB23}. Zolezzi et al. analyze the flow of users between these communities \cite{zolezzi2023characterizing}. However, their approaches do not incorporate sentiment analysis to determine whether these changes reflect only shifts in interaction patterns or also signify changes in political perspectives. Thus, the comprehensive understanding of opinion dynamic patterns of individuals switching from one political community to another remains relatively unexplored in the existing literature, to the best of our knowledge.

To address these gaps, we first curate a Twitter dataset with 2 million tweets about US president Donald Trump during the 2020 US presidential election year. We identify communities within two distinct time periods in the retweet interaction network. We propose a method to automatically detect users who change their political community. Then, we examine the salient features between users who shift their community and those who remain within their original community. We found that individuals who switch communities differ both in the topological aspects of their interaction patterns as well in the bias of their content towards Donald Trump.

The main contributions of this work are:
\begin{itemize}
\item We propose a machine learning framework that effectively identifies and characterize users who undergo online community changes. We find significant correlations between community-shifting and changes in sentiment towards Donald Trump.
\item We found that users who change their community have lower pagerank values, indicating reduced relevance in the online debate and their messages did not spread. A possible interpretation of these results is that a user changes community when they do not have strong affinities with their community and their messages have no response.
\end{itemize}

The paper is organized as follows. In the ``Data" section, we describe the data used in the study and how we collect it. In the ``Method" section, we describe the unsupervised community detection framework, the sentiment analysis pipeline, the definition of community-shifting users and how we characterize the users interactions. In the ``Results" section, we present the main findings of this work. Finally, we interpret these results in the ``Conclusions" section. 

\section{Data}

Twitter provided developers with a range of APIs to facilitate the development of applications. The Streaming API, which enabled real-time acquisition of tweet samples from the platform, allowed developers to filter the data based on language, terms, hashtags, and other criteria \cite{morstatter2013sample}. The dataset used in this study includes attributes such as tweet ID, text, timestamp, user ID, username. In the case of retweets, information about the original tweet's user account is also included.

For the purpose of our research, we used a dataset of tweets mentioning Donald Trump in the United States during the 2020 presidential election year. The dataset was collected using the keyword ``realDonaldTrump", the official Twitter account of President Donald Trump. The data collection process encompassed two distinct time intervals: May 9th to May 16th (hereafter referred as $t_1$) and June 10th to June 16th, 2020 (hereafter referred as $t_2$). A subset of the dataset employed in this study has been previously used in \cite{DBLP:conf/amw/AlbaneseFLB23}. To ensure consistency and alignment in the analysis, we follow the methodology of that work, considering only users who received or made more than five retweets in each time period. This threshold of five retweets serves as a trade-off between having a very low filter, which could compromise the robustness of community detection, and setting a very high filter, which could result in the exclusion of a significant amount of data. 

\subsection{Ethical Considerations and Data Availability}
This study strictly adheres to ethical considerations by exclusively utilizing publicly available data \cite{barakat2022community}. No data from private user accounts were collected or utilized. 
The tweet IDs were made publicly available for reproducibility and transparency purposes \footnote{https://github.com/fedealbanese/community-changing-users/}.

\section{Methods}

\subsection{Community detection}

A graph community refers to a subset of nodes within a larger network that exhibits a higher density of connections among its constituent nodes compared to nodes outside the community \cite{newman2004finding}. In the context of our Twitter dataset, it can be represented as a group of users who are more interconnected through retweets \cite{albanese2021improved}, for instance.

The interaction among individuals in our study is represented as a graph, denoted as $G = (N,E)$, where nodes ($N$) correspond to users and edges ($E$) represent retweets between them. To accurately capture the dynamics of multiple retweets from one user to another, we model the graph as a directed and weighted network. The edge points from the content creator to the retweeter, representing the flow of information through the network, originating from the content creator and spreading to the users who amplify the message \cite{DBLP:conf/amw/AlbaneseFLB23}.

The Stochastic Block Model (SBM) is a method that can be used for uncovering the latent communities within a network \cite{holland1983stochastic}. In contrast to alternative and popular methods that maximize modularity \cite{newman2004finding}, the SBM incorporates hierarchical priors and Bayesian inference techniques, enabling the model to provide a robust representation of the underlying data \cite{peixoto2019bayesian}. On the other hand, models that focus on maximizing modularity do not adequately account for deviations from the null model in a statistically consistent manner \cite{fortunato2016community}. Therefore, such models primarily serve as descriptive tools and lack the ability to support inferential conclusions. In this study, we employ the SBM implemented in the graph-tool python library to discern communities within directed and weighted networks \cite{peixoto2014hierarchical}. 

\subsection{Sentiment Analysis}

Sentiment analysis is a natural language processing technique employed to ascertain the sentiment or emotional tone expressed in a text. There are multiple approaches in order to classify the sentiment of a text such as Rule-Based algorithms \cite{qiu2010dasa}, machine learning models like Support Vector Machine (SVM) or Naive Bayes (NB) \cite{bai2011predicting} and deep learning networks \cite{zhang2018deep}. Considering the characteristics of our dataset (tweets with short and informal text), we use VADER (Valence Aware Dictionary and sEntiment Reasoner), that is a lexicon and rule-based sentiment analysis tool which was specifically design for social media content \cite{hutto2014vader}. For each tweet, VADER gives a sentiment score. According to the tool documentation, the score takes values from $1$ to $-1$ where a score bigger than $0.05$ is considered positive, a score between $0.05$ and $-0.05$ neutral, and lower than $-0.05$ negative. The sentiment score of a user is defined as the average sentiment of their tweets.

\subsection{Users interactions}

By considering users as nodes and retweets as edges within the retweet network, we characterize users by computing the following graph metrics:
\begin{enumerate}
    \item In-degree: the number of retweet the user did.
    \item Out-degree: the number of times they were retweeted.
    \item Pagerank: the importance of the user within the network.
    \item Betweenness centrality: the influence the user has over the flow of information in the network.
\end{enumerate}

These metrics portray the users relationship with the other users and distinguish the relevance in the network that they might have.

\section{Results}

\subsection{Communities in the retweet networks}

We use SBM to identify the communities of users of the retweet networks. In consideration of the bipartisan nature of US politics, characterized by the presence of two predominant parties (Democrats and Republicans) \cite{klein2020we,albanese2020analyzing}, we configured the SBM parameters to identify two communities within the networks.

\begin{figure}[h]
\centering     
\includegraphics[width=0.95\columnwidth]{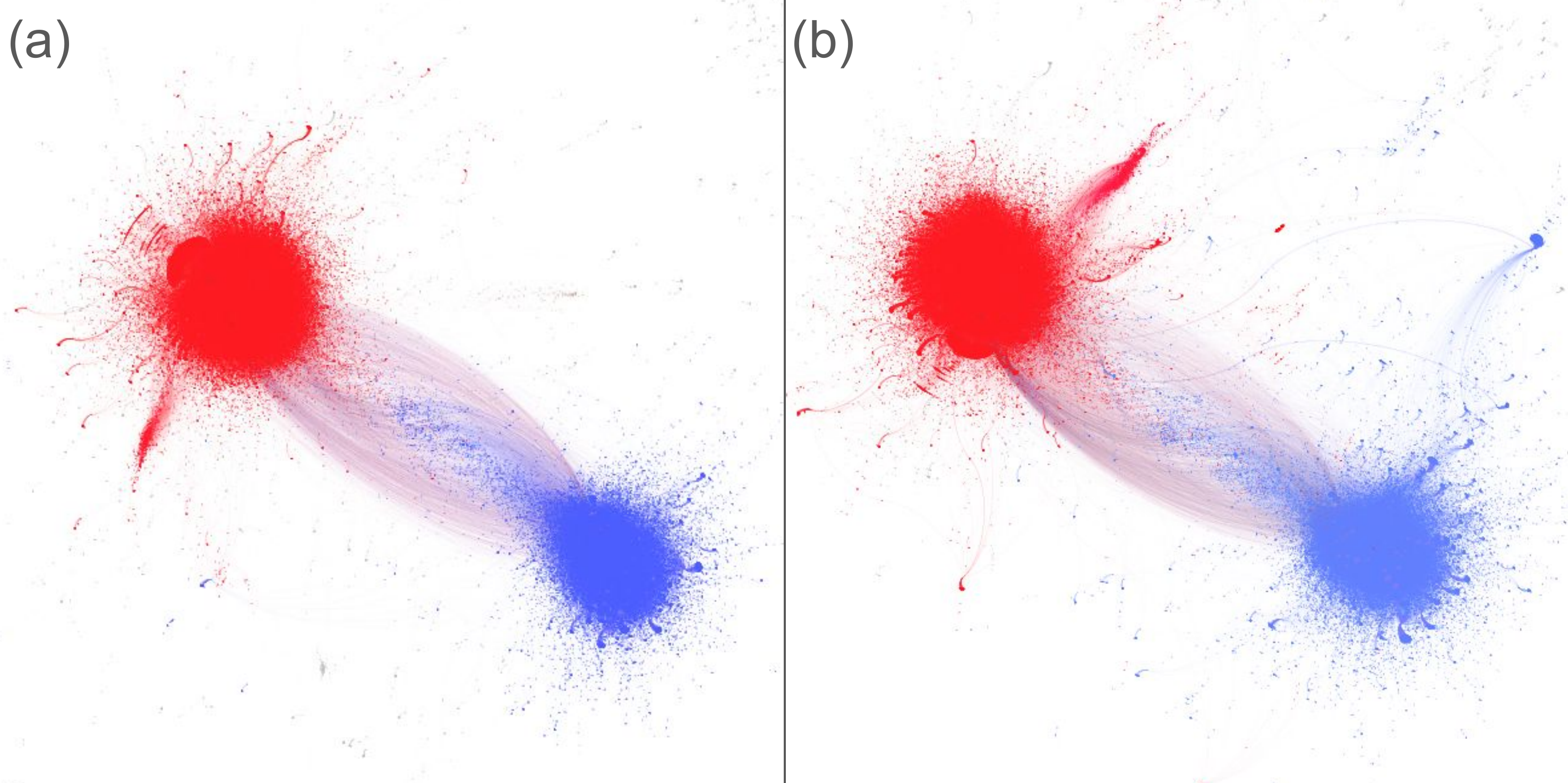}
\caption{Retweet network of the first time period from May 9th to May 16th, $t_1$, (a) and second time period from June 10th to June 16th, $t_2$, (b) during 2020 US presidential election. Each node is a Twitter user and each edge (directed and weighted) represents the retweets between two given users. The nodes are colored depending on its community: the ``republican community" users are in red and the ``democratic community" users are in blue.}
\label{figure1}
\end{figure}

In Fig.~\ref{figure1}, we visualize the retweet network and the results of the community detection algorithm for each time period. The users are concentrated in two communities, portraying the political polarization in that country. The graph visualizations were produced with Force Atlas 2 layout using Gephi software \cite{jacomy2014forceatlas2}. 

We would assign the labels ``republican community" to the community that have the users ``$@$realDonaldTrump"  and ``$@$Mike\_Pence" (the official Twitter accounts of the 2020 republican presidential and vice-presidential candidates) and ``democratic community" to the community that have the users ``$@$JoeBiden"  and ``$@$KamalaHarris" (the official accounts of the 2020 democratic presidential and vice-presidential candidates). The sentiment analysis results discussed in the next section further support this label assignment.

In table \ref{table_communities} we show main features of community analysis: the size and percentage of both groups at both times and how similar they are using the Jaccard Index. First, we should notice that republican communities are consistently bigger that the democratic ones. This imbalance can be attributed to the fact that data was collected using the   keyword ``realDonaldTrump", the Republican President of the United State at those times. Second, we can see most of users remain in both groups when they are compared at these two moments: $95.5\%$ for republicans and $92.7\%$ for democrats. Those who don't are the focus of our research in this work.

\begin{table}
\caption{\label{table_communities}The total number of users of each community for both time periods ($t_1$ and $t_2$).}
\begin{tabular*}{\textwidth}{@{}l*{15}{@{\extracolsep{0pt plus
12pt}}l}}
\br
& Republican community & Democratic community  \\
\mr
        \#users at $t_1$     & 12569 ($64.1\%$) & 7044 ($35.9\%$) \\
        \#users at $t_2$          & 11945 ($63.8\%$) & 6769  ($36.1\%$) \\
        Jaccard Index        & 0.955 & 0.927   \\
\br
\end{tabular*}
\end{table}

\subsection{Sentiment of the communities} 

We hypothesize that when users from different political communities mention Donald Trump, their expressed sentiments will exhibit notable discrepancies. We would expect that the republican community would have a more positive sentiment when mentioning Donald Trump and the democratic community would have a more negative sentiment when mentioning Donald Trump.

\begin{figure}[h]
\centering     
\includegraphics[width=0.95\columnwidth]{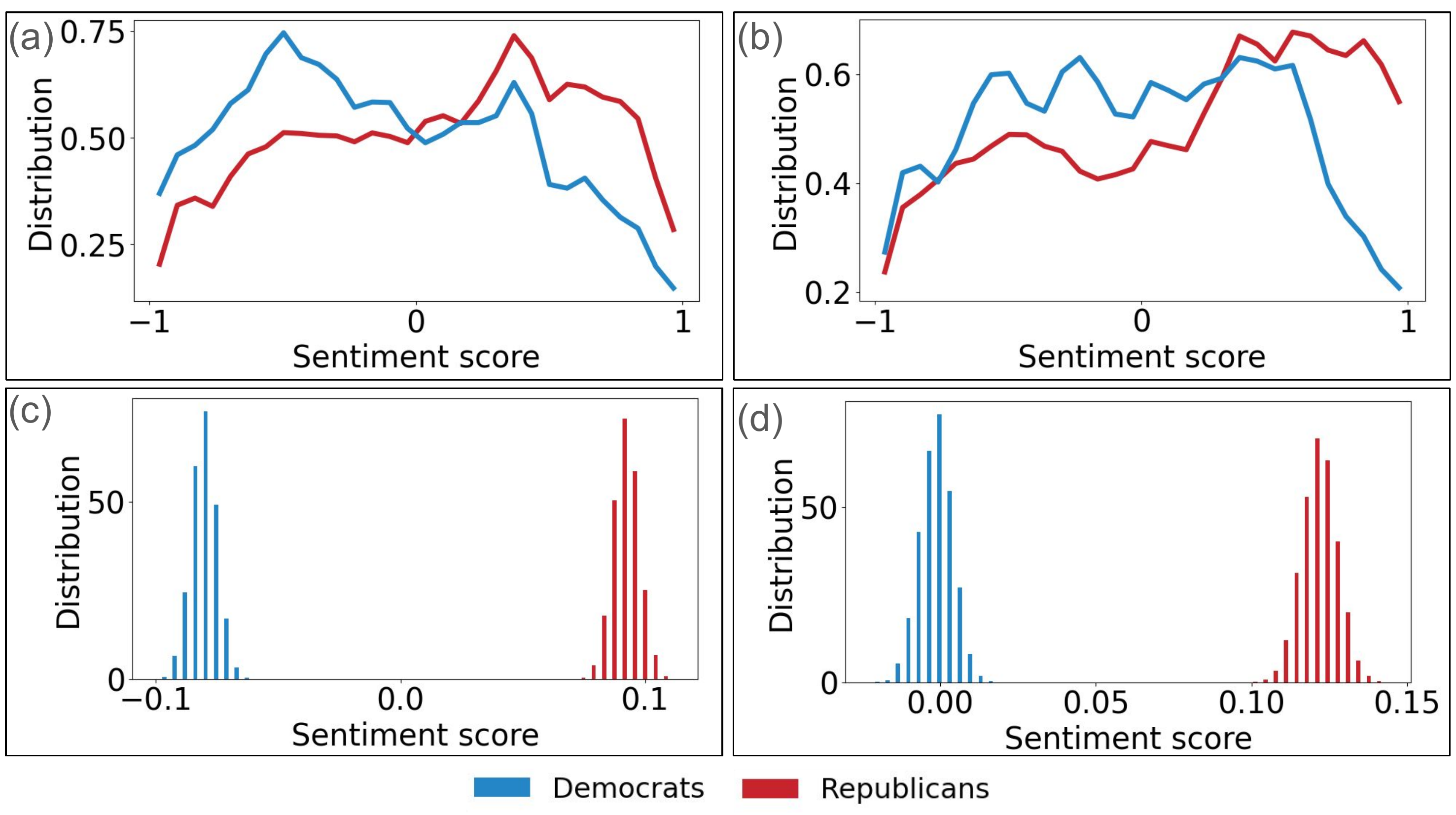}
\caption{Sentiment score of tweets belonging to the republican and the democratic community for the first time period from May 9th to May 16th, $t_1$, (a) and second time period from June 10th to June 16th, $t_2$, (b) during 2020 US presidential election. (c) and (d) show the Bootsrap sample of mean values at the same times.}
\label{figure2}
\end{figure}

In Fig. \ref{figure2}, we show the statistical distribution of the sentiment values of the users when mentioning President Donald Trump in each community and time periods. Given the complex nature of the data, we use bootstrap sampling, a method in which we repetitively draw sample data with replacement form the data distribution to estimate mean values and confidence intervals which gives simple representation of the data \cite{hesterberg2011bootstrap}. Fig. \ref{figure2} depicts both the original distribution of sentiment values for $t_1$ (a) and $t_2$ (b) and the bootstrap sample of mean values, also for $t_1$ (c) and $t_2$ (d). Results show that the users of the republican community have positive mean sentiment scores, meanwhile users from the democratic community have negative mean sentiment scores confirming our hypothesis. These result is consistent for both time periods. We also found that there are statistically significant differences between the sentiment scores of the communities ($p-value < 0.01$) using the Mann-Whitney U non-parametric test \cite{mann1947test} and the Kruskal-Wallis H-test \cite{kruskal1952use} at $t_1$ and at $t_2$. Also, it is worth noticing that the differences between the communities is stable thought time. 

While our analysis has revealed distinct sentiment scores for each community when mentioning Trump, not all community members align with these sentiments. In particular, $71.6\%$ of the users in the democratic community users express negativity, and $67.7\%$ of the users in the republican  community users have a positive sentiment score. Although these percentages represent a majority sentiment alignment, it's noteworthy that there exist users within these communities whose sentiments deviate from the prevailing trend.

\subsection{Users who change their community}

In this subsection, we aim to explore the differences between those users who change communities when comparing them at two different time instants $t_1$ and $t_2$. Understanding the differences between them can provide insights  on how individuals engage with other users from diverse ideological communities.

Formally, we define a user as a community-shifting user ($u_c$) when their community at time $t_1$ and $t_2$ are different. On the other hand,  we define a user as a non community-shifting user ($u_{nc}$) when their community at time $t_1$ and $t_2$ are the same.

\subsubsection{Topological analysis}

From the community detection performed previously, we can identify those users who change their community. Table \ref{table_graph_metrics} shows the graph metrics at $t_1$ for the users who change their community ($u_c$) and those than remains in their own ($u_{nc}$). Taking into account that $u_{nc}$ represent between $5\%$ and $8\%$ of the total members of each communities, mean values and confidence intervals were estimated through a sub-sampling approach. Specifically, we conducted $10.000$ iterations of random sub-sampling with replacement, measuring the mean of each sub-sample. The resulting distribution of means was then utilized to report the mean and standard deviation values.

\begin{table}
\caption{\label{table_graph_metrics}Mean graph features of users who change their community ($u_c$) and users who have the same community over time ($u_{nc}$) at $t_1$. Differences are statistically significant (Mann-Whitney U nonparametric test and Kruskal-Wallis H-test).}
\begin{tabular*}{\textwidth}{@{}l*{15}{@{\extracolsep{0pt plus
12pt}}l}}
\br
        & $u_c$ & $u_{nc}$  \\
\mr
        Indegree     &  $1.87 \pm 0.92$ & $6.19 \pm 18.40$ \\
        Outdegree        & $1.83 \pm 0.09$ & $6.13 \pm 0.28$  \\
        Pagerank        & $2.81e-06 \pm 7.73e-07$ & $6.30e-06 \pm 1.62e-6$  \\
        Betweenness centrality        & $7.88e-08 \pm 7.46e-08$ & $5.04e-07 \pm 5.80e-07$  \\
\br
\end{tabular*}
\end{table}

We found that, in all four metrics, there are statistically significant differences ($p-value < 0.01$)  between $u_c$ and $u_{nc}$ using the Mann-Whitney U nonparametric test \cite{mann1947test} and the Kruskal-Wallis H-test \cite{kruskal1952use}. It should be notices that $u_c$ have smaller values than $u_{nc}$ in all cases. These  behavioral patterns can be attributed to several factors. Firstly, users who will switch communities display lower values of in-degree and out-degree, meaning that they are not active retweeters nor retweeted. Also, the fact that $u_c$ had statistically lower pagerank values mean that those users were less relevant to the tweeter conversation and their messages did not spread in their original community. A possible interpretation of these results is that a user changes community when they do not have strong affinities with their community and their messages have no response.

\subsubsection{Sentiment score}

In this section we first analyze the mean sentiment scores towards Donald Trump for users from Republican and Democratic communities who either maintained ($u_{nc}$) or changed their community ($u_{c}$) affiliations at $t_1$.  We would like to know if the differences  between these two groups found in topological properties have a counterpart in this feature of tweets content.

Equivalently to table \ref{table_graph_metrics}, we report the mean values and confidence intervals using the same sub-sampling approach.
For Republican users, the mean sentiment score at $t_1$ for those who remained within their community was $s_{nc} = 0.081 \pm 0.433$, while for those who changed communities, it was $s_c =0.080 \pm 0.401$. Similarly, Democratic users who retained their community affiliation exhibited a mean sentiment score of $s_{nc} =-0.102 \pm 0.408$, whereas those who underwent a community change showed a mean score of $s_c =-0.087 \pm 0.456$.
Conducting Mann-Whitney U nonparametric test and the Kruskal-Wallis H-test statistical tests, we determined that the observed differences in sentiment scores between users who changed and those who did not were not statistically significant, which is consistent with the fact that the confidence intervals exceed the mean values in an order of magnitude. Therefore there is limited predictive capability based on sentiment scores, in contrast to the topological features.

On the other hand, we also characterize the variation of the average sentiment score between the first ($t_1$) and second ($t_2$) time periods of the users when mentioning President Donald Trump in each community (the sentiment score at $t_2$ minus the sentiment score at $t_1$). 

\begin{table}
\caption{\label{table_sentiment_metrics}Delta of the average sentiment score between the users of the republican and democratic community at $t_1$ and $t_2$ ($sentiment_{t_2} - sentiment_{t_1}$. Note: Differences are statistically significant (Mann-Whitney U nonparametric test and Kruskal-Wallis H-test).}
\begin{tabular*}{\textwidth}{@{}l*{15}{@{\extracolsep{0pt plus
12pt}}l}}
\br
        & Democrat at $t_1$ & Republican at $t_1$  \\
\mr
        Democrat at $t_2$     & $0.075 \pm 0.085$ & $-0.066 \pm 0.074$ \\
        Republican at $t_2$        & $-0.036 \pm 0.088$ & $0.035 \pm 0.082$   \\
\br
\end{tabular*}
\end{table}

Table \ref{table_sentiment_metrics} show the  variation in average sentiment  scores between $t_1$ and $t_2$  for the users of the republican and democratic community. 
Results show that republican users that maintain their community (those who were in the republican community at $t_1$ and remain there at $t_2$) change their sentiment score on average $0.035 \pm 0.082$. But users who were in the republican community at $t_1$ and change their community to  the democratic one at $t_2$ show variation in their average sentiment score  of $-0.066 \pm 0.074$. An equivalent effect was measure for democrat users. Therefore, we observe that users who change their communities also change their sentiment scores different than the user who maintain their community. 

These variations in the average sentiment score of the users who change their community are statistically significantly different ($p-value < 0.01$) than those of the users who did not change their community, according to the Mann-Whitney U nonparametric test and the Kruskal-Wallis H-test. The results show that the classification of users who undergo community changes aligns with significant changes in sentiment when mentioning Donald Trump between the first and second time periods. Despite the limited predictive capability of the sentiment scores, results demonstrate a relationship between changing the community and changing their sentiment towards Donald Trump.

\section{Conclusions}

In conclusion, our research has provided insights into the dynamics of online community changes and their implications for user behavior within social networks. Through the introduction of a machine learning framework, we have identified and characterized users undergoing community shifts.

Analyzing the changes in sentiment expressed by users during community-shifts can provide valuable insights into how these transitions could impact individuals' political views. Importantly, we have found significant correlations between community shifts and alterations in sentiment towards Donald Trump, shedding light on the interplay between ideological shifts and online interactions. 

Furthermore, our findings have showed that users who undergo community changes exhibit significantly lower pagerank values, suggesting a small relevance in the online discussion and limited message propagation thought the network. One plausible interpretation of these findings suggests that users may transition to a different community when their affinity to their current community weakens, and their messages have little resonance within their peers.

This work present an analysis of sentiment in tweets and user interactions, leveraging topological network metrics. The results contributes to advancing our understanding of the dynamics of online social networks and provides computational tools for effectively identifying and studying users who undergo community changes.

Future research involve the analysis of diverse datasets across various contexts, countries, and social networks. Also, examining an extended dataset spanning multiple years, employing different time windows, could offer valuable insights.


\end{document}